\documentclass[12pt,final,twocolumns]{elsart}

  \usepackage{latexsym,bm,amsmath,amssymb,amsfonts}
  \usepackage{epsfig,graphics,graphicx}
\usepackage{latexsym,bm,amsmath,amssymb,amsfonts}

\newcommand{\n}{{\bm n}}

\newcommand{\tr}{{\rm tr}  }

\newcommand{\beq}{\begin{eqnarray}}
\newcommand{\eeq}{\end{eqnarray}}

\begin{document}

\begin{flushright}

NIKHEF-2015-025\\
YITP15-59\\
%SFB/CPP-13-25\\
%\vspace{5mm}
\end{flushright}

\begin{frontmatter}

\parbox[]{16.0cm}{ \begin{center}
\title{Hemisphere jet mass distribution at finite $N_c$}

\author{Yoshikazu Hagiwara$^{\rm a}$, Yoshitaka Hatta$^{\rm b}$ and Takahiro Ueda$^{\rm c}$ }

%\address{$^{\rm a}$  Faculty of Pure and Applied Sciences, University
%of Tsukuba, Tsukuba, Ibaraki 305-8571, Japan}
\address{$^{\rm a}$ Department of Physics, Kyoto University, Kyoto 606-8502, Japan}

\address{$^{\rm b}$ Yukawa Institute for Theoretical Physics, Kyoto University, Kyoto 606-8502, Japan}
\address{$^{\rm c}$ Nikhef Theory Group, Science Park 105, 1098 XG Amsterdam, The Netherlands }
\end{center}

%\date{\today}

\begin{abstract}
We perform the leading logarithmic resummation of nonglobal logarithms for the single-hemisphere jet mass distribution in $e^+e^-$ annihilation including the finite-$N_c$ corrections. The result is compared with the previous all-order result in the large-$N_c$ limit as well as fixed-order perturbative calculations.
\end{abstract}
}

\end{frontmatter}

\section{Introduction}

The hemisphere jet mass distribution is an event shape variable in $e^+e^-$-annihilation defined as the distribution of invariant mass  in a single hemisphere whose axis coincides with the thrust axis. As is usually the case with all event shapes, it receives logarithmically enhanced perturbative corrections when the shape variable becomes small. However, unlike other event shapes for which systematic resummation methods are available (see \cite{Hoang:2014wka,Banfi:2014sua} and references therein), the resummation of logarithms for the hemisphere jet mass distribution has turned out to be  thorny and so far remained unsatisfactory even at the leading logarithmic level. This is due to the presence of the so-called nonglobal logarithms \cite{Dasgupta:2001sh}
   which arise from the energy-ordered radiation of soft gluons in a restricted region of phase space.

The difficulty of resumming nonglobal logarithms stems from the fact that one has to keep track of the distribution of an arbitrary number of secondary soft gluons emitted  at large angle. (For a recent review, see \cite{Luisoni:2015xha}.) The original work by Dasgupta and Salam \cite{Dasgupta:2001sh} employed a Monte Carlo algorithm, valid  to leading logarithmic accuracy and in the large-$N_c$ approximation, to actually generate soft gluon cascades on a computer. Later, Banfi, Marchesini and Smye (BMS) \cite{Banfi:2002hw} reduced the problem, still at large-$N_c$, to solving a nonlinear integro-differential equation.
This latter approach paved the way for the inclusion of the finite-$N_c$ corrections in the resummation \cite{Weigert:2003mm} which has been recently put on a firmer ground  \cite{Caron-Huot:2015bja},  and the first quantitative finite-$N_c$ result can be found in \cite{Hatta:2013iba}. However, so far only one particular observable (`interjet energy flow') has been computed and the full impact of the finite-$N_c$ resummation is yet to be uncovered.

In this work, we apply the method developed in \cite{Hatta:2013iba} to the hemisphere jet mass distribution and numerically carry out the resummation of nonglobal logarithms at finite-$N_c$, thereby achieving the full leading logarithmic accuracy for this observable.\footnote{Recently, the resummation of nonglobal logarithms to next-to-leading logarithmic order has  been discussed \cite{Caron-Huot:2015bja,Larkoski:2015zka}. In particular, Ref.~\cite{Caron-Huot:2015bja} explicitly derived the full NLL evolution kernel at finite $N_c$. See, also, an earlier suggestion \cite{Avsar:2009yb} that the NLL resummation of nonglobal logarithms should be related to the NLL BFKL resummation via a conformal transformation.}
 In the next section we define the observable and introduce the BMS equation which resums the nonglobal logarithms in the large-$N_c$ limit. In Section 3, we discuss the resummation strategy at finite-$N_c$. It turns out that a naive application of the previous method is plagued by large numerical errors, and we shall propose a refined method which cures this problem. In Section 4, we present the numerical result and compare it with the previous all-order result at large-$N_c$  \cite{Dasgupta:2001sh} as well as the recent fixed-order calculations \cite{Schwartz:2014wha,Khelifa-Kerfa:2015mma}.

\section{Hemisphere jet mass distribution}

 Consider a two-jet event in $e^+e^-$-annihilation with the center-of-mass energy $Q$.
Without loss of generality, we assume that the quark jet is right-moving with momentum $p^\mu=\frac{Q}{2}(1,0,0,1)\equiv \frac{Q}{2}(1,{\bm n}_R)$ and the antiquark jet is left-moving with $\bar{p}^\mu=\frac{Q}{2}(1,0,0,-1)\equiv \frac{Q}{2}(1,{\bm n}_L)$. Suppose soft gluons with momentum $k_i^\mu=\omega_i (1, {\bm n}_i)$ (${\bm n}=(\sin \theta \cos \phi, \sin \theta \sin \phi, \cos\theta)$) are emitted in the right hemisphere $0\le \theta_i \le \frac{\pi}{2}$.
 The invariant mass in the right hemisphere is
 \beq
M_R^2=\left(p+\sum_i k_i\right)^2\approx \sum_i 2p\cdot k_i =\sum_i Q\omega_i (1-\cos \theta_i)\,.
 \eeq
 We shall be interested in the probability
 \beq
 P_{LR}(\rho)=\frac{1}{\sigma}\int^\rho_0 \frac{d\sigma}{d\rho'}d\rho'\,,
  \eeq
 that the rescaled invariant mass
 \beq
 \rho'=\frac{M_R^2}{Q^2}=\sum_i \frac{\omega_i (1-\cos \theta_i)}{Q}\,,
 \eeq
  is less than some value $\rho<1$.
When $\rho \ll 1$, one has to resum large logarithms $\ln^n 1/\rho$ in the perturbative calculation of $P_{LR}(\rho)$. As observed in \cite{Dasgupta:2001sh}, this resummation consists of two parts. One is the Sudakov double logarithms $(\alpha_s \ln^2 1/\rho)^n$ which can be resummed via exponentiation. The other is the nonglobal logarithms $(\alpha_s \ln 1/\rho)^n$ which arise from the fact that measurement is done only in a part of phase space (i.e., in the right hemisphere). The latter resummation affects various single-hemisphere event shapes in $e^+e^-$ annihilation and DIS \cite{Dasgupta:2001sh}. It is also relevant to the so-called soft function in the dijet mass distribution $d^2\sigma/dM_RdM_L$ in the asymmetric limit $M_L\gg M_R$ \cite{Kelley:2011ng,Hornig:2011iu,Schwartz:2014wha}.

So far, the resummation of nonglobal logarithms for $P_{LR}$  has been carried out only in the large-$N_c$ limit \cite{Dasgupta:2001sh}, or to finite orders of perturbation theory at large-$N_c$
\cite{Schwartz:2014wha} and finite-$N_c$ \cite{Khelifa-Kerfa:2015mma}. As stated in the introduction, we shall perform the all-order leading logarithmic resummation  at finite-$N_c$ along the lines of \cite{Weigert:2003mm,Hatta:2013iba}.
To explain our approach, it is best to start with the BMS equation which resums both the Sudakov and nonglobal logarithms in the large-$N_c$ limit \cite{Banfi:2002hw}. Adapted to the hemisphere jet mass distribution \cite{Schwartz:2014wha}, the equation reads
\beq
 \partial_\tau  P_{\alpha\beta}=  N_c\int \frac{d\Omega_\gamma}{4\pi}
 {\mathcal M}_{\alpha\beta}(\gamma)
\Bigl( \Theta_{L}(\gamma) P_{\alpha \gamma} P_{\gamma\beta}
-P_{\alpha\beta} \Bigr)
\,,   \label{bms}
\eeq
 where
 \beq
  {\mathcal M}_{\alpha\beta}(\gamma) =\frac{1-\cos \theta_{\alpha\beta}}{(1-\cos \theta_{\alpha \gamma})(1-\cos \theta_{\gamma\beta})}\,, \label{soft}
  \eeq
  is the soft gluon emission kernel and we defined
 \beq
 \tau=\frac{ \alpha_s}{\pi}\ln \frac{1}{\rho}\,.
 \eeq
 $\Theta_{L/R}(\gamma)$ is the `step function' which restricts the angular integral $d\Omega_\gamma=d\cos \theta_\gamma d\phi_\gamma$ to the left/right hemisphere. [Below we also use a shorthand notation $\int_{\scriptsize{L/R}} d\Omega$ to represent this.]
 In (\ref{bms}), $P_{\alpha\beta}=P(\Omega_\alpha, \Omega_\beta)$ is the generalization of $P_{LR}$ defined for arbitrary pairs of solid angle directions.

Taken at its face value, Eq.~(\ref{bms}) is ill-defined. When $\alpha$ or $\beta$ is in the right hemisphere, the $d\Omega_\gamma$ integral in the second term on the right-hand-side (the virtual term) is divergent, and this is precisely the situation  $(\alpha\beta)=(LR)$ we are eventually interested in. Physically, this collinear divergence should be cut off by the kinematical effect, yielding the Sudakov factor $e^{-{\mathcal O}(\alpha_s) \ln^2 1/\rho}$. However, since the Sudakov factor is well understood anyway, one can leave it out of consideration by defining
  \beq
 P_{\alpha \beta}\equiv \exp\left(-2C_F \tau \int_R\frac{d\Omega_\gamma}{4\pi} {\mathcal M}_{\alpha \beta}(\gamma)
 \right)  g_{\alpha \beta}\,. \label{su}
\eeq
($C_F=\frac{N_c^2-1}{2N_c}$ and $2C_F\approx N_c$ in the large-$N_c$ limit.)
Unlike (\ref{bms}), the equation satisfied by $g_{\alpha\beta}$ is well-defined and amenable to analytical and numerical approaches. In particular, Ref.~\cite{Schwartz:2014wha} analytically calculated $g_{LR}$ to five loops using the   hidden SL(2,${\mathbb R}$) symmetry of the BMS equation \cite{Hatta:2009nd}.

\section{Resummation at finite $N_c$}

We now turn to the physically relevant case $N_c=3$.
Temporarily forgetting about the issue of the collinear divergence, we recapitulate the resummation strategy developed in \cite{Blaizot:2002np,Weigert:2003mm,Hatta:2013iba}.
First we make the formal identification
\beq
P_{\alpha\beta} \leftrightarrow \frac{1}{N_c} \mbox{tr}\, U_\alpha U^\dagger_\beta \,, \label{id}
\eeq
   where $U_\alpha$ is the Wilson line in the fundamental representation of SU($N_c$) from the origin to infinity in the $\Omega_\alpha$ direction. The product in (\ref{id}) represents the propagation of the $q\bar{q}$ jets (`dipole') in the eikonal approximation.
  As $\tau$ is increased, more and more soft gluons are emitted from the dipole and also from the secondary gluons. This can be simulated as a stochastic process in which the Wilson lines receive random kicks in the color SU($N_c$) space, and is described by the following Langevin equation in discretized `time' $\tau$ \cite{Hatta:2013iba}\footnote{We write the evolution (\ref{alt}) in a slightly different, but equivalent form compared to Ref.~\cite{Hatta:2013iba}.  It should be understood that  various exponentials are meaningful only to ${\mathcal O}(\epsilon)$ \cite{Hatta:2013iba}, although in practice we  keep all orders in $\sqrt{\epsilon}$ in order to preserve the unitarity of  $U_\alpha$.}
\beq
U_\alpha(\tau+\varepsilon) &=& e^{iS^{(2)}_\alpha} e^{iA_\alpha} U_\alpha (\tau) e^{iB_\alpha}e^{iS_\alpha^{(1)}}\,, \label{alt}
\eeq
 where
 \beq
 S^{(i)}_\alpha = \sqrt{\frac{\varepsilon}{4\pi}} \int_R d\Omega_\gamma  \frac{(\n_\alpha -\n_\gamma)^k}{1-\n_\alpha \cdot \n_\gamma} \, t^a \xi^{(i)k}_{\gamma a}\,, \qquad (i=1,2) \label{s1}
 \eeq
% \beq
%  S_\alpha=\sqrt{\frac{2\varepsilon}{4\pi}} \int_R d\Omega_\gamma  \frac{(\n_\alpha -\n_\gamma)^k}{1-\n_\alpha \cdot \n_\gamma} \, t^a \xi^{(1)k}_{\gamma a} \label{s2}
% \eeq
% \beq
 %A_\alpha = \sqrt{\frac{\varepsilon}{4\pi}}\int d\Omega_\gamma \frac{(\n_\alpha -\n_\gamma)^k}{1-\n_\alpha \cdot \n_\gamma} \left(-\Theta_L(\gamma) U_\gamma t^a U_\gamma^\dagger  \xi^{(1)k}_{\gamma a} +\Theta_{R}(\gamma)\, t^a  \xi^{(2)k}_{\gamma a} \right)\,, \label{lan}
% \eeq
 \beq
 A_\alpha = -\sqrt{\frac{\varepsilon}{4\pi}}\int_L d\Omega_\gamma \frac{(\n_\alpha -\n_\gamma)^k}{1-\n_\alpha \cdot \n_\gamma} U_\gamma t^a U_\gamma^\dagger  \xi^{(1)k}_{\gamma a}\,, \label{lan}
 \eeq
 \beq
 B_\alpha= \sqrt{\frac{\varepsilon}{4\pi}} \int_L d\Omega_\gamma  \frac{(\n_\alpha -\n_\gamma)^k}{1-\n_\alpha \cdot \n_\gamma} \, t^a \xi^{(1)k}_{\gamma a}\,, \label{lan2}
 \eeq
 and $\xi^{(1)}$, $\xi^{(2)}$ are the Gaussian noises
 \beq
 \langle \xi^{(i)k}_{\gamma a}(\tau) \xi^{(j)l}_{\gamma' b}(\tau')\rangle =\delta^{ij}\delta_{\tau,\tau'} \delta (\Omega_\gamma-\Omega_{\gamma'})\delta_{ab}\delta^{kl}\,. \label{noise}
 \eeq
This is equivalent to the following `Fokker-Planck' equation to be compared with (\ref{bms})
\beq
\!\!\partial_\tau \langle P_{\alpha\beta}\rangle_\xi
% &=& N_c\int \frac{d\Omega_\gamma}{4\pi} {\mathcal M}_{\alpha\beta}(\gamma) \left\{\Theta_L(\gamma)\left(\langle P_{\alpha \gamma}P_{\gamma\beta}\rangle_\xi -\frac{\langle P_{\alpha\beta}\rangle_\xi}{N_c^2}\right)
%-\frac{2C_F}{N_c}\langle P_{\alpha\beta}\rangle_\xi \right\} \nonumber \\
= \! N_c\int \frac{d\Omega_\gamma}{4\pi} {\mathcal M}_{\alpha\beta}(\gamma) \left\{\Theta_L(\gamma)\bigl(\langle P_{\alpha \gamma}P_{\gamma\beta}\rangle_\xi -\langle P_{\alpha\beta}\rangle_\xi\bigr)\!
-\frac{2C_F}{N_c}\Theta_R(\gamma)\langle P_{\alpha\beta}\rangle_\xi \right\},
\label{evolve}
\eeq
 where $\langle...\rangle_\xi$ denotes averaging over the noises.
In principle, $P_{LR}(\tau)$ at finite-$N_c$ can be evaluated by  computing $\frac{1}{N_c}\mbox{tr}U_L U^\dagger_{R}$ for a given random walk trajectory  with the initial condition $U_\alpha(\tau=0)=1$, and then averaging over many trajectories. In this calculation, it suffices to define $U_\alpha$ in the left hemisphere and at a single point $\alpha=R$ in the right hemisphere.

However, this strategy does not apply straightforwardly to our present problem.  $\langle P_{LR}\rangle_\xi$ quickly goes to zero due to the collinear divergence in the Sudakov factor.\footnote{This problem was not encountered in \cite{Hatta:2013iba} because there $\alpha$ and $\beta$ were always confined in the unobserved part of phase space.} One may try to regularize the divergence by introducing a cutoff $\delta$ and extract the finite part
  \beq
\langle g_{LR}(\tau)\rangle_\xi = \lim_{\delta\to 0} \ \exp\left(2C_F \tau \int_R\frac{d\Omega_\gamma}{4\pi} {\mathcal M}_{LR}(\gamma)
 \right)_{\delta}  \langle P^\delta_{LR}(\tau)\rangle_\xi\,.
\eeq
Unfortunately, this does not work in practice because $\langle P^\delta_{LR}\rangle_\xi$ becomes very small and the exponential factor becomes very large as $\delta \to 0$. It is difficult to numerically achieve the precise cancelation between the two factors.

As a matter of fact, the same problem was already noticed  in the original Monte Carlo simulation at large-$N_c$ \cite{Dasgupta:2001sh}. There the authors subtracted the Sudakov contribution step-by-step, by modifying the emission probability as ${\mathcal M}_{\alpha \beta}(\gamma) \to {\mathcal M}_{\alpha \beta}(\gamma)-\Theta_R(\gamma){\mathcal M}_{LR}(\gamma)$.
Here  we shall implement a similar subtraction directly in the evolution of $U_\alpha$. The origin of the collinear divergence can be traced to the factors $e^{iS^{(i)}_\alpha}$ in (\ref{alt}). They give,  after averaging over the noise,
%angular integrals $\int d\Omega_\gamma$ in (\ref{lan}) and (\ref{lan2}) over the right hemisphere
%\beq
%\exp\left(i \sqrt{\frac{\varepsilon}{4\pi}} \int_R d\Omega_\gamma  \frac{(\n_\alpha -\n_\gamma)^k}{1-\n_\alpha \cdot \n_\gamma} \, t^a \xi^{(j)k}_{\gamma a}\right)\,, \qquad (j=1,2) \label{mat}
%\eeq
% which can be factored out from $e^{iA_\alpha}$ and $e^{iB_\alpha}$ (i.e., $e^{a+b}\approx e^a e^b$).\footnote{These exponentials can be factored out  because the noises with different indices or at different time steps are uncorrelated (\ref{noise}). By the same reason, the order of exponentials in the product in (\ref{check})  does not matter. It should be understood that  various exponentials are meaningful only to ${\mathcal O}(\epsilon)$ \cite{Hatta:2013iba}, although in practice we  keep all orders in $\sqrt{\epsilon}$ in order to preserve the unitarity of  $U_\alpha$.}
\beq
&&\!\!\!\!\!\!\langle e^{iS^{(2)}_\alpha}e^{iS_\alpha^{(1)}}e^{-iS_\beta^{(1)}}e^{-iS^{(2)}_\beta}\rangle_\xi\nonumber \\
 &&\!\!\!\!\!\!\!\!\!= \prod_{i}^{1,2}\left\langle \exp\!\left(i \sqrt{\frac{\varepsilon}{4\pi}} \int_R d\Omega_\gamma  \frac{(\n_\alpha -\n_\gamma)^k}{1-\n_\alpha \cdot \n_\gamma} \, t^a \xi^{(i)k}_{\gamma a}\right) \!\exp\!\left(\!-i \sqrt{\frac{\varepsilon}{4\pi}} \int_R d\Omega_\gamma  \frac{(\n_\beta -\n_\gamma)^k}{1-\n_\beta \cdot \n_\gamma} \, t^a \xi^{(i)k}_{\gamma a}\right)
\right\rangle_{\!\xi}  \nonumber \\
&&\!\!\!\!\!\!= \exp\left(-2C_F\varepsilon \int_R \frac{d\Omega_\gamma}{4\pi} {\mathcal M}_{\alpha\beta}(\gamma)\right)\,, \label{check}
\eeq
 which is indeed the Sudakov factor (\ref{su}) generated in a single step.
 (\ref{check}) can be checked by using the identity \cite{Hatta:2013iba}
 \beq
 \!\!\!\!{\mathcal M}_{\alpha\beta}(\gamma)=2{\mathcal K}_{\alpha\beta}(\gamma)-{\mathcal K}_{\alpha\alpha}(\gamma)-{\mathcal K}_{\beta\beta}(\gamma)\,, \quad {\mathcal K}_{\alpha\beta}(\gamma)\equiv \frac{(\n_\alpha-\n_\gamma)\cdot (\n_\gamma-\n_\beta)}{2(1-\n_\alpha \cdot \n_\gamma)(1-\n_\gamma\cdot \n_\beta)}\,.
 \eeq
 In the special case $(\alpha\beta)=(LR)$, we have that ${\mathcal K}_{LR}(\gamma)\equiv 0$  and
 \beq
 \int_R \frac{d\Omega_\gamma}{4\pi} {\mathcal M}_{LR}(\gamma)= \int_R \frac{d\Omega_\gamma}{4\pi} \bigl(-{\mathcal K}_{LL}(\gamma)-{\mathcal K}_{RR}(\gamma) \bigr)
 = \frac{\ln 2}{2} + \int_R \frac{d\Omega_\gamma}{4\pi} \frac{1}{1-\n_R\cdot \n_\gamma}\,. \label{ln}
 \eeq
  We see that the singularity entirely comes from the second order term in the expansion of $e^{iS^{(i)}_R}$.

It is thus tempting to remove the factor $e^{iS^{(i)}_\alpha}$ altogether and use a modified evolution equation $\widetilde{U}_\alpha(\tau+\varepsilon)= e^{iA_\alpha} \widetilde{U}_\alpha(\tau) e^{iB_\alpha}$. However,  this also removes an essential part of the nonglobal logarithms. The reason is that the linear term in the expansion of $e^{iS_\alpha(\xi)}=1+iS_\alpha(\xi)+\cdots$ can give   finite contributions when the Gaussian noise $\xi$ is contracted with that implicit in $U_\gamma t^a U_\gamma^\dagger$ in (\ref{lan}). Physically, the factor $U_\gamma t^a U^\dagger_\gamma = U_A^{ab}t^b$ (with $U_A$ being the Wilson line in the adjoint representation) represents the emission of real gluons which is restricted to the left hemisphere. These gluons together with the original $q\bar{q}$ pair  form a QCD antenna which coherently emits the softest gluon in the right hemisphere, thereby producing nonglobal logarithms.

 To make the last statement more concrete, we follow the evolution (\ref{alt}) analytically up to $\tau=2\varepsilon$ (two steps) and collect the non-Sudakov contributions. We find
\beq
\!\!\!\! \!\!\langle g_{LR}(\tau)\rangle_\xi&\sim& 2C_F N_c\tau^2\int_L \frac{d\Omega_\gamma}{4\pi} \int_R \frac{d\Omega_\lambda}{4\pi}
\bigl({\mathcal K}_{LL}(\gamma)+{\mathcal K}_{RR}(\gamma)\bigr) \left({\mathcal K}_{L\gamma}(\lambda) + {\mathcal K}_{\gamma R}(\lambda)-{\mathcal K}_{\gamma\gamma}(\lambda)\right) \nonumber \\
&=&-C_F N_c\tau^2\int_L \frac{d\Omega_\gamma}{4\pi} \int_R \frac{d\Omega_\lambda}{4\pi} {\mathcal M}_{LR}(\gamma) \left( {\mathcal M}_{L\gamma}(\lambda)+{\mathcal M}_{\gamma R}(\lambda)-{\mathcal M}_{LR}(\lambda)\right) \nonumber \\
&=& -\pi^2\frac{C_F N_c\tau^2}{12}\,, \label{mis}
\eeq
  in agreement with the lowest order (two-loop) result \cite{Dasgupta:2001sh,Schwartz:2014wha}.\footnote{At this order, we have to interpret  $2\varepsilon^2=\tau(\tau-\varepsilon)\approx \tau^2$ to correct an error in iteratively solving a discretized differential equation.}
In the first line of (\ref{mis}), the factors ${\mathcal K}_{L\gamma }$ and ${\mathcal K}_{\gamma R}$ come from the linear terms in $e^{iS_L}\approx 1+iS_L$ and $e^{iS_R}\approx 1+iS_R$, respectively. They both seem to be essential for obtaining the correct result.

Importantly, however,  the term ${\mathcal K}_{\gamma R}(\lambda)$ vanishes when integrating over the azimuthal angle $\phi_\lambda$
\beq
\int_0^{2\pi} d\phi_\lambda {\mathcal K}_{\gamma R}(\lambda) = \int_0^{2\pi} d\phi_{\lambda\gamma} \frac{\cos \theta_\lambda -1-\cos \theta_\gamma + \cos \theta_{\lambda\gamma}}{2(1-\cos \theta_\lambda)(1-\cos \theta_{\lambda\gamma})}=0\,,
\eeq
 where we used $\cos \theta_\lambda >0>\cos \theta_\gamma$. Moreover, by following the evolution (\ref{alt}) a few more steps, it is easy to convince oneself that the linear term $iS_R$ does not produce nonglobal logarithms to all orders because this term always reduces to factors like ${\mathcal K}_{\gamma^{(n)}R}(\lambda)$ (after contracting with the $n$-th gluon emitted in the left hemisphere) and vanishes when  integrating over $\phi_\lambda$ in the right hemisphere.
 This observation brings in a major simplification in our resummation strategy. We can  neglect the factors $e^{iS^{(1,2)}_R}$ in (\ref{alt}) for $\alpha=R$ and use the modified Langevin equation
\beq
\widetilde{U}_R(\tau+\epsilon)=e^{iA_R} \widetilde{U}_R (\tau) e^{iB_R}\,. \label{fin}
\eeq
 As for $U_\alpha$ in the left hemisphere, we may continue to use the same evolution (\ref{alt}). Actually,  we can make a slight improvement which speeds up the numerical simulation. The two independent noises $\xi^{(1,2)}$ defined in the right hemisphere always give identical contributions for the observable at hand. Therefore, we can eliminate one of them and use a modified equation
 \beq
 U_\alpha(\tau+\varepsilon)=  \exp\left(i\sqrt{\frac{2\varepsilon}{4\pi}} \int_R d\Omega_\gamma  \frac{(\n_\alpha -\n_\gamma)^k}{1-\n_\alpha \cdot \n_\gamma} \, t^a \xi^{(1)k}_{\gamma a}\right) e^{iA_\alpha}U_\alpha(\tau)e^{iB_{\alpha}}\,. \label{new}
 \eeq
 Note the factor of $\sqrt{2}$. One can check that (\ref{new}) leads to the same equation (\ref{evolve})  for the product of two Wilson lines.\footnote{We have checked numerically that (\ref{new}) and (\ref{alt}) give equivalent results. The equivalence may not hold for more complicated observables.}
 Using these Langevin equations, we finally compute the average
 \beq
 \langle g_{LR}(\tau)\rangle_\xi= e^{\tau C_F \ln 2 } \frac{1}{N_c}\langle \mbox{tr} \, U_L(\tau)\widetilde{U}_R^\dagger(\tau)\rangle_\xi\,. \label{final}
 \eeq
The multiplicative factor in front subtracts the finite part of the Sudakov factor (\ref{ln}) which is included in the evolution of $U_L$.

\section{Numerical results and discussions}

\begin{figure}[tbp]
  \includegraphics[width=160mm]{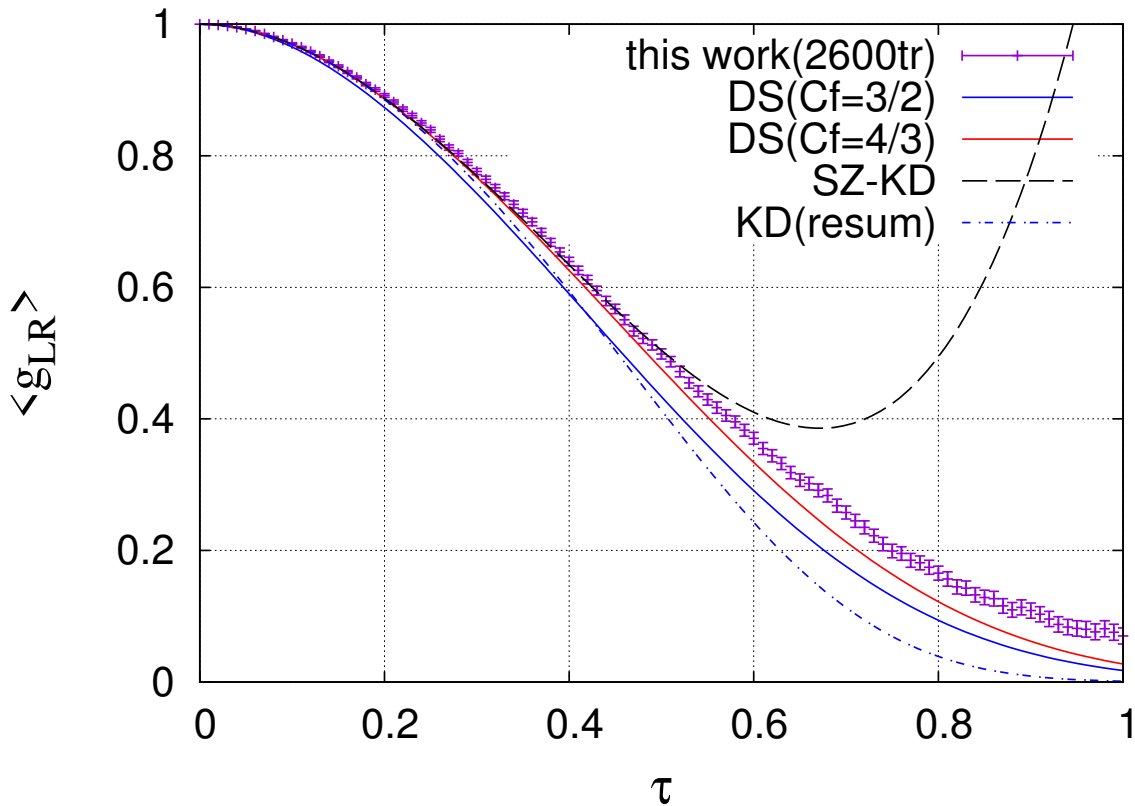}
 \caption{$\langle g_{LR}(\tau)\rangle$ at $N_c=3$ as a function of $\tau$ obtained by averaging over 2600 random walks.  The error bars indicate the standard error. Data points are plotted every $0.01/\varepsilon=200$ random walk steps. Various curves are explained in the text.  \label{fig1}}
\end{figure}

The numerical procedure is explained in Ref.~\cite{Hatta:2013iba} which we refer to for details.
We discretize the solid angle $1 \ge \cos \theta \ge -1$ and $2\pi >\phi\ge 0$ into a lattice of $80\times 80$ grid points and put  a SU(3) matrix $U_\alpha$  at each grid point on the left hemisphere $\cos \theta<0$.  In addition, we define a SU(3) matrix $\widetilde{U}_R$ at a single point ${\bm n}_R$ in the right hemisphere.  The Gaussian noise $\xi$ is randomly generated at all grid points and at each time step.\footnote{In \cite{Hatta:2013iba}, the authors inadvertently assumed that the noise $\xi$ (at each time step) is independent of $\phi$  at the degenerate points $\cos\theta=\pm 1$. Fortunately,  this was innocuous for the observable considered in \cite{Hatta:2013iba}. However, for the present observable this causes a  systematic error in the evaluation of the Sudakov integral (the first term of (\ref{ln})) already at small-$\tau$ because the integration region $\int d\Omega_R$ includes the point $\cos\theta=1$ (which was not the case in \cite{Hatta:2013iba}). In the present simulations, we fixed this problem by generating $\xi$ at different values of $\phi$ independently also at $\cos\theta=\pm 1$.
%The previous numerical result in \cite{Hatta:2013iba} is little affected by this modification, thanks to the fact that the relevant Sudakov integral does not include the points $\cos \theta=\pm 1$.
}
We then evolve $U_\alpha$ and $\widetilde{U}_R$ according to (\ref{new}) and (\ref{fin}), respectively, with $\varepsilon=5\times 10^{-5}$ and the initial conditions  $U_\alpha=\widetilde{U}_R=1$. As in the previous work \cite{Hatta:2013iba}, we observe large event-by-event fluctuations. In order to obtain a reasonably smooth curve, we typically have to average over ${\mathcal O}(10^3)$ random walks. Fig.~\ref{fig1} shows the result from 2600 random walks.\footnote{We also performed simulations with different discretization parameters ($160\times 40$ and $80\times 40$ lattices, and  $\varepsilon=10^{-4}$) and found that the results are consistent with each other.} In the same figure, we make comparisons with the following results in the literature: The blue line is a parameterization of the all-order Monte Carlo result in the large-$N_c$ limit by Dasgupta and Salam (DS) \cite{Dasgupta:2001sh}
\beq
g^{DS}(\tau)=\exp\left(-C_F N_c \frac{\pi^2 \tau^2}{12}\frac{1+(a \tau/2)^2}{1+(b\tau/2)^{c}}\right)\,, \label{ds}
\eeq
with  $a=0.85N_c$, $b=0.86N_c$ and $c=1.33$. Here we set $C_F\approx N_c/2=1.5$ which is what was actually used in \cite{Dasgupta:2001sh}.
 The black dashed line is a combination of the fixed-order analytical results by Schwartz and Zhu (SZ) to five-loop at large-$N_c$ \cite{Schwartz:2014wha} and  Khelifa-Kerfa and Delenda (KD) to four-loop at finite $N_c$ \cite{Khelifa-Kerfa:2015mma}
%\beq
%g^{SZ-KD}(\tau)&=&1-C_FN_c\frac{\pi^2}{12}\tau^2 + \frac{C_F N_c^2\zeta_3}{6}\tau^3
%-\frac{1}{24}\left(\frac{25}{8}C_F N_c^3\zeta_4 -\frac{13}{5}C_F^2 N_c^2\zeta_2^2\right) \tau^4 \nonumber \\
%&& \qquad+\left(-\frac{\pi^2\zeta_3}{360}+\frac{17}{480} \zeta_5\right)N_c^5\tau^5\,,
%\label{sz}
% \eeq
 \beq
 g^{SZ-KD}(\tau)&=&1-C_FN_c\frac{\pi^2}{12}\tau^2 + \frac{C_F N_c^2\zeta_3}{6}\tau^3
-\frac{1}{24}\left(\frac{25}{8}C_F N_c^3\zeta_4 -\frac{13}{5}C_F^2 N_c^2\zeta_2^2\right) \tau^4 \nonumber \\
&& \qquad+\frac{1}{120}\left(-8C_F^2 N_c^3\zeta_2 \zeta_3+\frac{17}{2}C_F N_c^4 \zeta_5\right)\tau^5\,,
\label{sz}
 \eeq
  where $C_F=4/3$. Actually, the complete finite-$N_c$ result at ${\mathcal O}(\tau^5)$ is not available, and the above formula is a well-motivated guess \cite{Khelifa-Kerfa:2015mma} which reduces to the known result in the large-$N_c$ limit. Finally, the blue dash-dotted line is the following `resummed' expression suggested by KD based on their four-loop result
\beq
\!\!\!g^{KD(resum)}(\tau)=\exp\left(-C_F N_c \frac{\pi^2\tau^2}{12}+\frac{C_FN_c^2\zeta_3\tau^3}{6} -\frac{\pi^4}{135}\left(\frac{25}{8}C_F N_c^3+C_F^2 N_c^2\right)\frac{\tau^4}{16}\right)\,, \label{kd}
\eeq
 with $C_F=4/3$. Note, however, that at the moment it is not known whether the nonglobal logarithms actually exponentiate to all orders.\\

We see that our result agrees very well with the most-advanced fixed-order result (\ref{sz}) up to $\tau \lesssim 0.5$. Beyond this, the perturbative result quickly deviates and eventually blows up.  It has been observed that higher loop contributions alternate in sign and converge rather poorly \cite{Rubin:2010fc}. In addition, fixed-order results are numerically sensitive to the $1/N_c$--suppressed  corrections when $\tau\sim {\mathcal O}(1)$. This can be partly remedied in the resummed formula (\ref{kd}).   On the other hand, the all-order large-$N_c$ result (\ref{ds})  stays close to our curve up to $\tau=1$. In fact, the difference can be partly accounted for by choosing  $C_F=4/3$ in (\ref{ds}), which is what was actually suggested by DS as the likely functional form at finite-$N_c$ and has been used for phenomenological purposes \cite{Dasgupta:2002dc,Dasgupta:2012hg}. This is shown by the red line in Fig.~\ref{fig1}. To correct the remaining difference, we independently determined $a,b,c$ in (\ref{ds}) with $C_F=4/3$ and obtained
 \beq
a=0.62\pm 0.06 \,, \qquad b=0.06 \pm 0.03 \,, \qquad c=0.37\pm 0.04 \,.
\eeq
\\

%extracted from the large-$N_c$ Monte Carlo simulations. Somewhat surprisingly, the agreement is better for (\ref{ds}) with $C_F=1.5$ than with $C_F=\frac{N_c^2-1}{2N_c}=\frac{4}{3}$, the latter being  The reason of this is unclear to us in view of the facts that we do have $C_F=\frac{4}{3}$ in (\ref{mis}) and various angular integrals are numerically evaluated to a few percent (or better) accuracy. On the other hand, the most-advanced fixed-order result (\ref{sz}) significantly deviates from the full result already when $\tau\sim 0.5$.
 % A better agreement is obtained if the perturbative series is cut off at ${\mathcal O}(\tau^4)$ (or at some even power of $\tau$) and exponentiated  as in (\ref{kd}). However, the result is not stable against the inclusion of the ${\mathcal O}(\tau^5)$ terms in the resummation .

In conclusion, we have completed the  resummation project for the single-hemisphere jet mass distribution initiated in \cite{Dasgupta:2001sh} by including the finite-$N_c$ corrections to all orders. We find that the finite-$N_c$ effect is numerically small, and this is consistent with the previous finding in \cite{Hatta:2013iba}. However, it should be kept in mind that the observables calculated at finite-$N_c$ so far are defined in $e^+e^-$ annihilation where the two outgoing jets are represented by the product of two Wilson lines $\mbox{tr}\, U_\alpha U_\beta^\dagger$. In hadron-hadron collisions, or in processes including hard gluons, one needs to consider the evolution of more complicated objects such as $\mbox{tr}(U_\alpha U_\beta^\dagger U_\gamma U_\delta^\dagger)$ and $\mbox{tr}(U_\alpha U_\beta^\dagger) \mbox{tr} (U_\gamma U_\delta^\dagger)$  (cf., \cite{Hatta:2013qj}). The finite-$N_c$ effects in the resummation of nonglobal logarithms for these observables have not been studied so far.

\section*{Acknowledgements}
The work of T.~U. is supported by the ERC Advanced Grant
  no. 320651, ``HEPGAME''. Numerical computations have been mostly carried out at the Yukawa Institute Computer Facility.

\end{document}